\documentclass[aps,prl,twocolumn,showpacs,refautoscript,reprint,longbibliography]{revtex4-2}
\usepackage{boldline,multirow}
\usepackage{float}
\usepackage{graphicx,natbib}
\usepackage{bm,amsmath,amssymb,wasysym,amsfonts,dcolumn}
\usepackage[colorlinks=true, urlcolor=blue, linkcolor=blue, citecolor=blue, pdfborder={0 0 0}]{hyperref}
\usepackage{ulem}
\usepackage[usenames,dvipsnames]{xcolor}
\usepackage{lipsum}
\usepackage{epstopdf}
\usepackage{cleveref}
\crefname{figure}{Fig.}{Figs}
\crefname{table}{Table}{Tables}
\crefname{section}{Sec.}{Sections}

\newcommand{\ket}[1]{\left| #1 \right>} 
\newcommand{\bra}[1]{\left< #1 \right|} 

\newcommand{\rom}[1]{\mathrm{#1}}

\newcommand{\lsf}{l_{\rom{sf}}}
\newcommand{\lof}{l_{\rom{of}}}

\begin{document}

\title{Orbital relaxation length from first-principles scattering calculations}
		
\author{Max Rang}
\author{Paul J. Kelly}
\affiliation{Faculty of Science and Technology and MESA$^+$ Institute for Nanotechnology, University of Twente, P.O. Box 217,
	7500 AE Enschede, The Netherlands}

\date{\today}

\begin{abstract}
The orbital Hall effect generates a current of orbital angular momentum perpendicular to a charge current.
Experiments suggest that this orbital current decays on a long length scale that is of the order of the spin flip diffusion length or longer. 
We examine this suggestion using first-principles quantum mechanical scattering calculations to study the decay of orbital currents injected from an orbitally-polarized lead into thermally disordered bulk systems of selected transition metals.
We find that the decay occurs over only a few atomic layers. On this length scale the orbital current may be converted into a spin current if the spin Hall angle is sufficiently large, as for Pt.
In Cu, Cr and V with small spin Hall angles, the conversion into a spin current is negligible  in the bulk and significant conversion only occurs at interfaces. 
\end{abstract}

\pacs{}

\maketitle

\section{Introduction}
The orbital analogue of the spin Hall effect (SHE) \cite{Dyakonov:pla71, Dyakonov:zetf71, Hirsch:prl99, Valenzuela:nat06}, the orbital Hall effect (OHE) \cite{Tanaka:prb08, Kontani:prl09, Go:prl18, Choi:nat23}, is predicted to have significantly higher conversion rates than the SHE, in particular for 3$d$ transition metals \cite{Jo:prb18, Salemi:prm22}.
Experimental results that appear consistent with this theoretical prediction \cite{Sala:prr22, Rothschild:prb22, Zhang:apl22, Lyalin:prl23} invoke long relaxation lengths for the orbital currents. 
For Ti, for example, relaxation lengths of 50-$60 \pm 15 \,$nm \cite{Choi:nat23} or $47\pm11 \,$nm \cite{Hayashi:cmp23} are extracted from magneto-optical Kerr effect (MOKE) and orbital Hall torque (OHT) experiments, respectively; 
for Cr, MOKE experiments are consistent with a relaxation length of $6.6 \pm 0.6 \,$nm \cite{Lyalin:prl23} while OHT experiments yield a value of $6.1\pm 1.7 \,$nm \cite{Lee:cmp21}. 
Such large values are at variance with the standard picture of orbital quenching in solids according to which orbitally nondegenerate states contain equal amounts of $+m$ and $-m$ character because of time-reversal symmetry \cite{Mohn:03, Tinkham:03}. 
Noting that all states in a solid are orbitally nondegenerate at room temperature because thermal disorder eliminates all rotational symmetry, we expect that an orbitally polarized state $|e^{+im\phi}\rangle$ injected into a material at room temperature can only hop to states containing equal amounts of $|e^{+im\phi}\rangle$ and $|e^{-im\phi}\rangle$ character so the orbital polarization is quenched on the length scale of the hopping. 

Although it has been argued in a number of recent publications that $l_{\rm of}$, the orbital analogue of the spin-flip diffusion length $l_{\rm sf}$, is short \cite{Salemi:prm21, Belashchenko:prb23, Urazhdin:prb23}, the methods used in these papers were not suitable for making quantitative estimates of $l_{\rm of}$, in particular in the presence of various types of disorder. 
It is the purpose of the present work to illuminate the region between theoretical studies on perfectly crystalline and defect-free bulk materials on the one hand and macroscopic experiments on inhomogeneous materials at room temperature on the other, by determining the orbital angular momentum (OAM) relaxation length using quantum mechanical scattering calculations for thermally disordered solids.

\section{Method}
\begin{figure}[b]
\centering
\includegraphics[width=8.4cm]{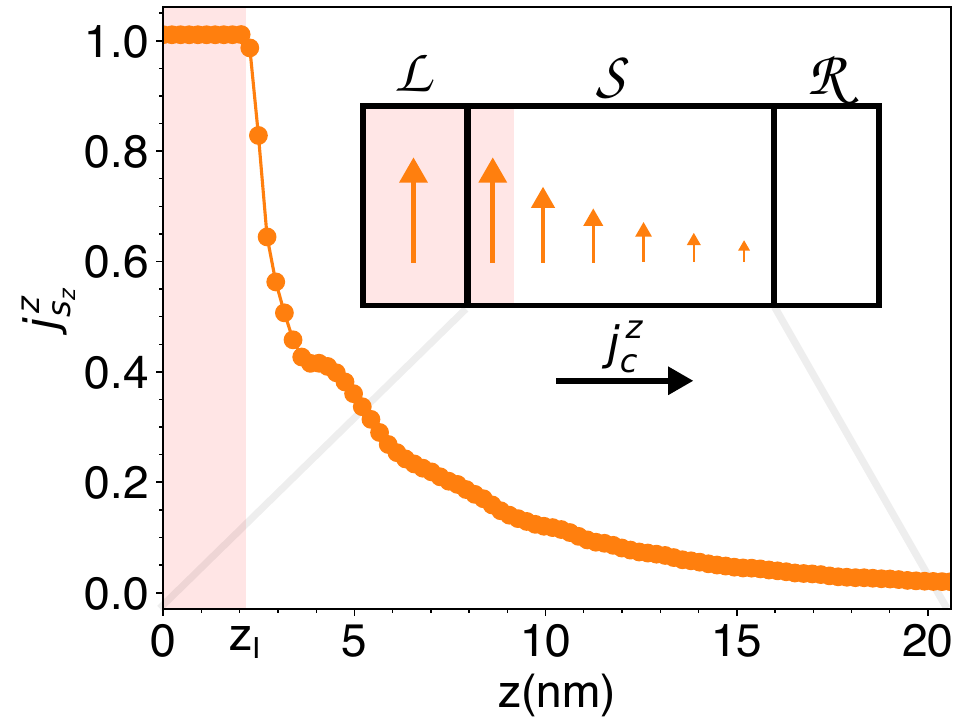}
\caption{Exponential decay of a spin-polarized current injected from a ballistic Pt lead into room-temperature (RT) thermally disordered Pt \cite{Wesselink:prb19}. To visualize the current in the lead, the interface between ballistic (pink) and diffusive Pt is displaced into the scattering region to $z=z_I$ \cite{footnote2}.  
}
\label{fig:lsf}
\end{figure}

The calculations involve large scattering geometries (thousands of atoms) and take temperature-induced disorder into account in the adiabatic approximation \cite{LiuY:prb15}.
The formalism implemented in the Twente Quantum Transport ({\sc tqt}) \cite{Xia:prb06, Starikov:prb18, Wesselink:prb19, Nair:prb21a} code considers the disordered material of interest as a scattering region ($\mathcal{S}$) sandwiched between semi-infinite crystalline left ($\mathcal{L}$) and right ($\mathcal{R}$) leads, see \Cref{fig:lsf}. 
Solution of the time-independent Schr\"odinger equation by matching the wavefunctions at the $\mathcal{L|S}$ and $\mathcal{S|R}$ interfaces takes the form of a set of inhomogeneous linear equations \cite{Ando:prb91}.
The resulting wave function contains all information on the linear response of the system to an external potential \cite{Khomyakov:prb05, Starikov:prb18}. 
On computing the solution in a basis of tight-binding linearized muffin-tin orbitals (TB-LMTO) \cite{Andersen:prl84, *Andersen:85, *Andersen:prb86}, we use the continuity equation to extract interatomic (spin and orbital) currents from the wave function \cite{Wesselink:prb19}. 
The flux $j$ of the property $\hat{X}$ between atoms $P$ and $Q$ is
\begin{equation}
\!\!\!j^{PQ}_X = \frac{1}{i\hbar}\left[\bra{\Psi_P} \hat{X}_P \hat{H}_{PQ} \ket{\psi_Q} - \bra{\Psi_Q} \hat{H}_{QP} \hat{X}_{Q} \ket{\Psi_Q}\right],
\end{equation}
where $\hat{X}=1$ leads to an expression for the charge current, $\hat{X}=\hat{S}_\alpha$ yields the spin current polarized in the $\alpha$ direction and $\hat{X}=\hat{L}_\alpha$ results in the orbital current polarized in the $\alpha$ direction, a current of orbital angular momentum \cite{footnote1}.
If the operator $\hat{X}$ is such that $\bra{\Psi_P} \hat{X} \ket{\Psi_Q} \ne 0$, this definition of the flux is not well defined since the continuity equation requires local conservation. 
In particular, we cannot use the definition of the OAM in the so-called ``modern theory of orbital polarization'' \cite{Thonhauser:prl05, Cysne:prb22, Pezo:prb22} to define a current in the scattering approach; instead, we use the atom-centered approximation of the OAM operators to compute OAM currents.
Interatomic currents were previously used to compute the spin Hall angle (SHA), spin-flip diffusion length, spin memory loss and the effect of interface disorder on those parameters \cite{WangL:prl16, Wesselink:prb19, Gupta:prl20, Nair:prl21, Gupta:prb21, Gupta:prb22, LiuRX:prb22}.

To construct a spin-polarized current, the (effective \cite{footnote2}) left lead is modified to shift the energy of one of the spin channels so that the Fermi surface becomes completely spin-polarized \cite{Wesselink:prb19}. 
A charge current injected from the left lead carries a spin current that decays exponentially in the scattering region of thermally disordered bulk diffusive Pt for $z>z_I$ as illustrated in \Cref{fig:lsf}. 
At room temperature, a value of the spin-flip diffusion length $\lsf\approx 5.2$ nm \cite{Wesselink:prb19, Nair:prl21} can be extracted directly. 
Constructing a spin-polarized system this way requires introducing a Zeeman term in the Hamiltonian for the left lead, albeit only in the spin degree of freedom
\begin{equation}
\hat{H} = \hat{B}\cdot \hat{S},
\label{eq2}
\end{equation}
where the amplitude of $\hat{B}$ is chosen to completely spin-polarize the Fermi surface. 

\begin{figure}[t]
\centering
\includegraphics[width=8.6cm]{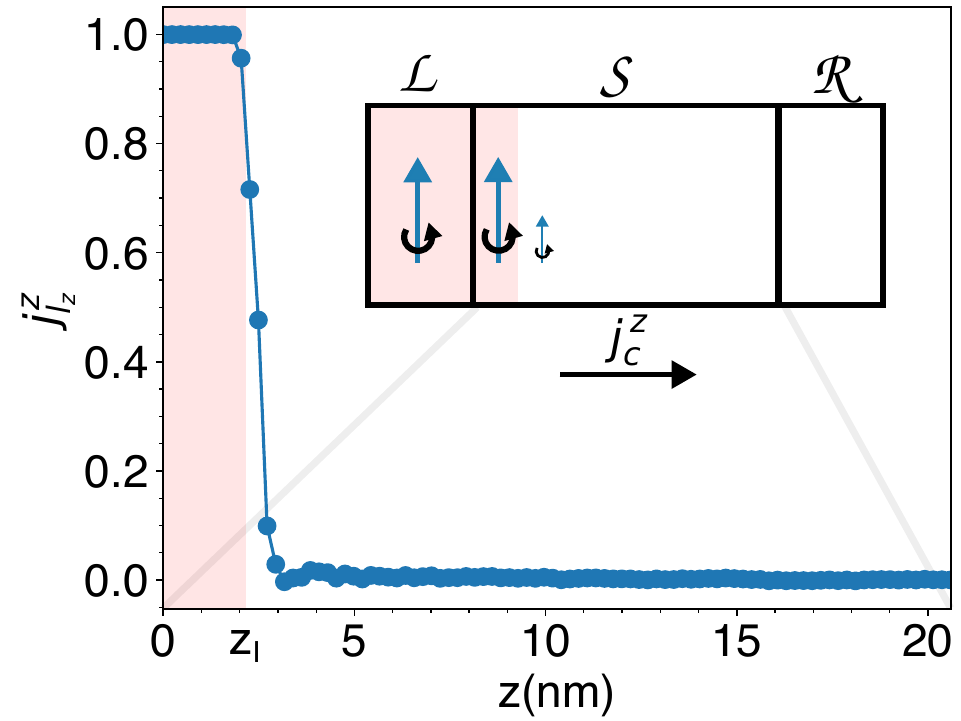}
\caption{An orbitally-polarized current injected from an orbitally-polarized effective left lead \cite{footnote2} decays on a length scale of only a few atomic layers in RT thermally disordered Pt when SOC is not included.}
\label{fig:lof}
\end{figure}

To compute an orbital relaxation length, we want to inject an orbitally polarized current.
To do so, we introduce an orbital Zeeman term into the Hamiltonian for the (effective) left lead
\begin{equation}
\hat{H} = \hat{B}\cdot \hat{L},
\label{eq3}
\end{equation}
where the choice of amplitude of $\hat{B}$ is not so straightforward as in the spin case. 
Unlike the spin degrees of freedom which are only coupled weakly through the spin-orbit coupling (SOC), modelled in our case as 
\begin{equation}
\hat{H}_{\rm so} = \xi \hat{L}\cdot \hat{S},
\label{eq4}
\end{equation}
the orbital degrees of freedom are much more strongly coupled by the crystal potential so that introducing a significant Zeeman orbital term completely distorts the electronic band structure.
The resulting computational setup is sketched in the inset to \Cref{fig:lof} and used to inject a current of OAM into thermally disordered, diffusive Pt.
The most striking feature of the figure is the very short decay length $l_{\rm of}$ for the OAM current. 

\section{Computational details}
The workflow for these calculations is as follows. We begin with a standard density functional theory (DFT) calculation for each transition metal system using the {\sc questaal} code \cite{Pashov:cpc20} with the von Barth-Hedin LDA functional \cite{vonBarth:jpc72} in combination with a TB-LMTO $spd$ basis and a 21$\times 21\times$21 $\mathbf{k}$-point grid.
The Kohn-Sham \cite{Kohn:pr65} atomic sphere \cite{Andersen:prl84, *Andersen:85, *Andersen:prb86} potential is extracted and used to construct the TB-LMTOs in the scattering calculation. 
The wave-function matching problem \cite{Ando:prb91}, which has been shown to be equivalent to the more conventional non-equilibrium Green's function (NEGF) method \cite{Khomyakov:prb05}, is set up and solved in the {\sc tqt} code. 
Periodic boundary conditions in the $x$ and $y$ directions ($z$ is the transport direction) require sampling the two-dimensional Brillouin zone with a sufficiently dense $\mathbf{k}$-point grid. From previous calculations \cite{Wesselink:prb19}, a grid equivalent to 160$\times$160 $\mathbf{k}$-points for a 1$\times$1 unit cell is more than sufficient. Here we have used a 29$\times$29 $\mathbf{k}$-point grid with a 5$\times$5 supercell, which is equivalent to 145$\times$145 for a 1$\times$1 unit cell.
Thermal disorder is modelled by randomly displacing the atoms from their equilibrium positions in the scattering region (for $z>z_I$) with a Gaussian distribution of displacements. We take multiple configurations of the random disorder and use the average to simulate the thermodynamic mean. 
In practice, 10 configurations is usually sufficient to obtain results uniquely distinguishable from the noise inherent in the random displacements.
The root mean square disorder parameter is iteratively chosen so that the resistivity of the system is as close to the experimental room temperature resistivity as desired. 
``Room temperature (RT)'' is defined to be $T=300\,$K.

\section{Results}
\begin{figure}[t]
\centering
\includegraphics[width=8.6cm]{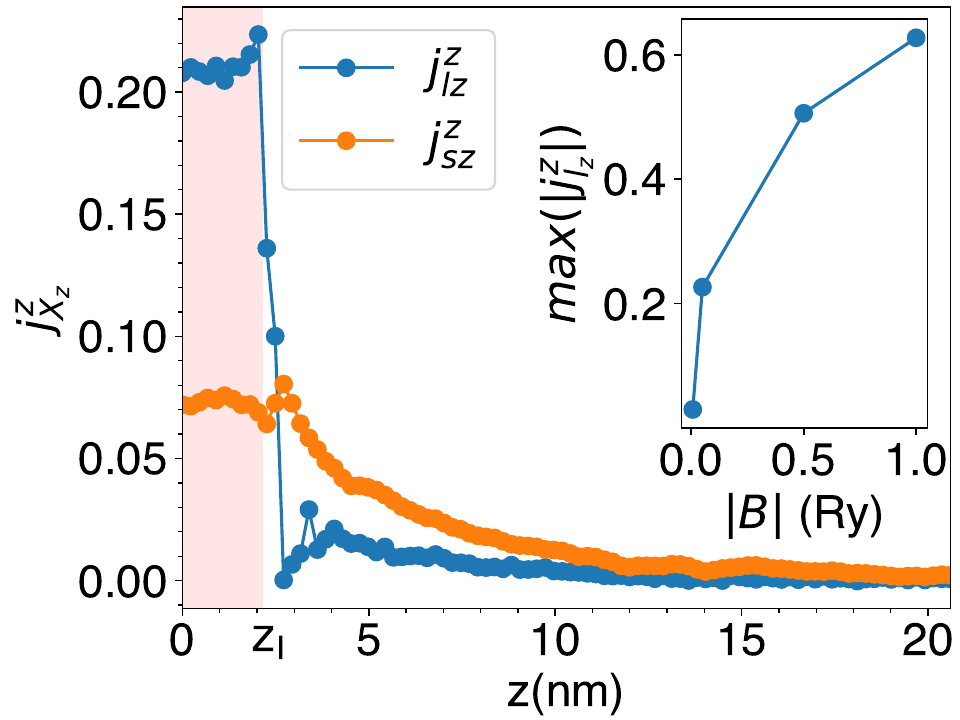}
\caption{A current of OAM $j_l$, orbitally-polarized in the $z$ direction ($j_{lz}$) and injected into RT thermally disordered Pt in the $z$ direction ($j_{lz}^z$), is converted into a spin current $j_s$ within a few atomic layers after which $j_s$ decays on a length scale consistent with the spin-flip diffusion length of Pt, $l_{\rm Pt}\equiv l_{\rm sf}^{\rm Pt}$. 
For the orbital Zeeman term in the lead, a value of $B=0.05\,$Ry was used.
The inset shows how the degree of orbital polarization depends on the size of the orbital Zeeman splitting in the lead.}
\label{fig:lof_ws}
\end{figure}

The orbital current injected into thermally disordered Pt decays very rapidly as seen in \Cref{fig:lof}. 
The orbital polarization is unity inside the (effective) left lead \cite{footnote2} but decreases to 2.5\% of this value within five atomic layers in diffusive Pt. Fitting an exponential curve to these results is not very meaningful and would result in an ``orbital diffusion length'' shorter than $1 \,$nm. 
On plotting the spin and orbital currents together in \Cref{fig:lof_ws} for a different, smaller value of $B=0.05\,$Ry with SOC switched on, it can be seen that the orbital current is converted into a spin current within a few layers in disordered Pt. In the few atomic layers where the orbital current decays rapidly, the spin current increases from around 0.06 to 0.08. The polarization of this spin current is not tremendously high, around 8\% at its peak. It decays on a length scale consistent with the spin-flip diffusion length $l_{\rm Pt}\equiv l_{\rm sf}^{\rm Pt} \approx 5.2\,$nm. 

\begin{figure}[t]
\centering
\includegraphics[width=8.6cm]{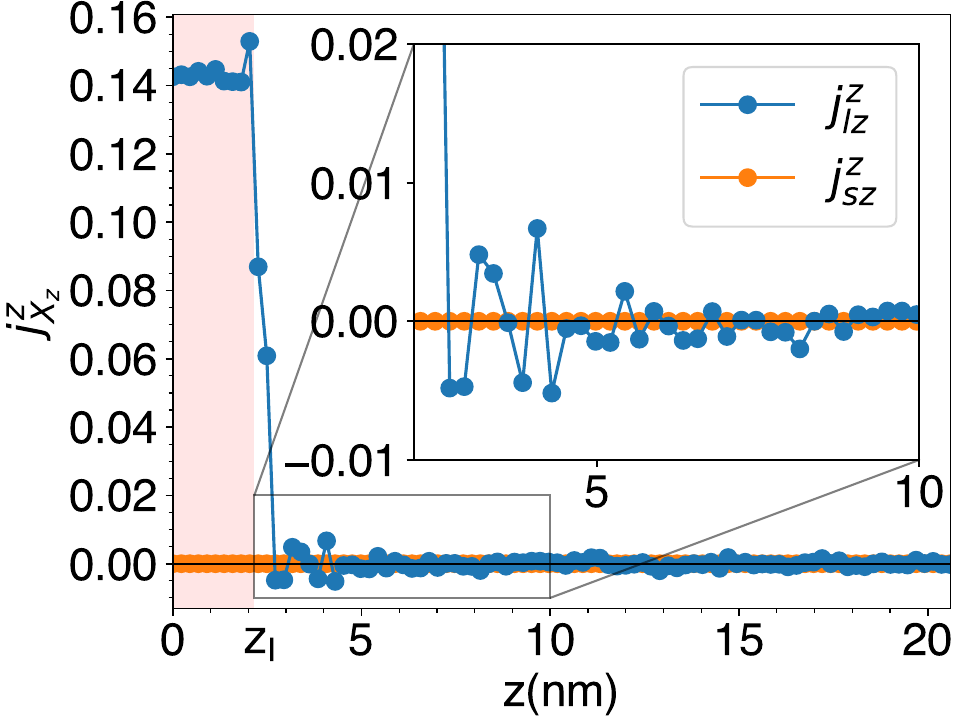}
\caption{Same as \Cref{fig:lof_ws} but with the SOC switched off. The orbitally-polarized current $j_l$ injected into RT thermally disordered Pt decays within a few atomic layers.}
\label{fig:lof_nosoc_ws}
\end{figure}

\begin{figure}[b]
\centering
\includegraphics[width=8.6cm]{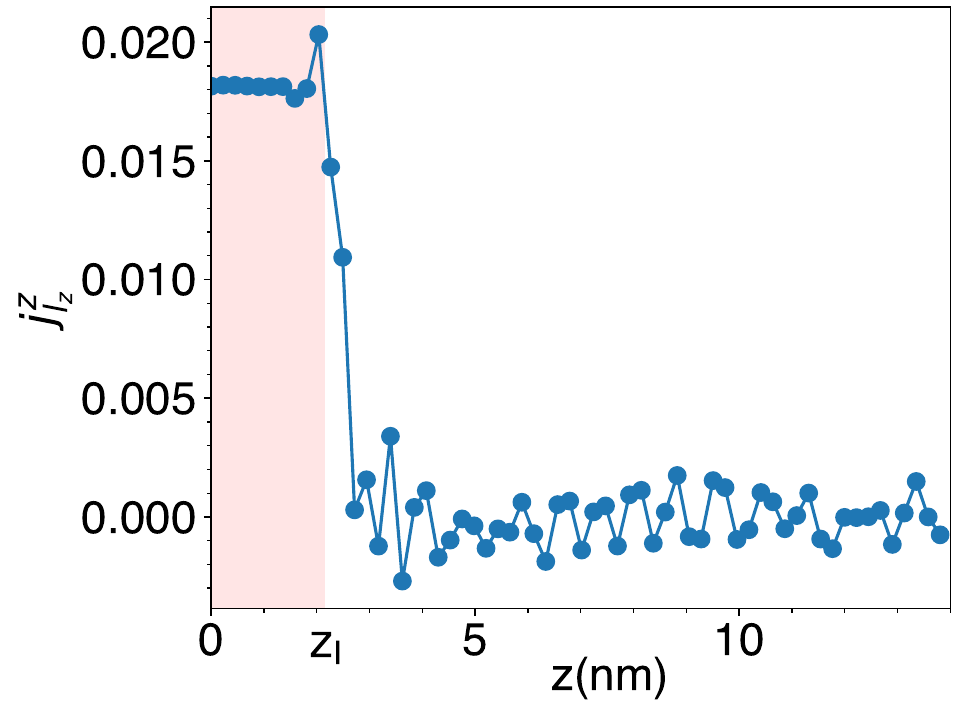}
\caption{An orbitally-polarized current generated with a value of $B=0.007\,$Ry in the left lead and injected into $T=0\,$K, ordered, Pt decays within a few atomic layers.}
\label{fig:lof_T0_noSOC_ws}
\end{figure}

The effect of the Pt SOC can be studied by switching it off in the scattering region by setting $\xi$ to zero in \eqref{eq4}, see \Cref{fig:lof_nosoc_ws}. 
The short decay length of the orbital current remains unchanged, of order a few atomic layers. The spin current is zero throughout the scattering region since in the absence of SOC no orbital current is converted into spin current. 
The sizable difference in the orbital polarization in the lead in the absence of SOC ($\sim14\%$ without versus $\sim21\%$ with SOC) comes from the slight change in the band structure when the SOC is turned off. 
This does not affect the principal result that $l_{\rm of}$ is very short, of the order of a few atomic layers. 
The same calculation but without thermal lattice disorder and with $B=0.007\,$Ry shows that disorder does not significantly affect $l_{\rm of}$, \Cref{fig:lof_T0_noSOC_ws} \cite{footnote3}. 
The fcc structure with T=0 has inversion symmetry while the snapshots of the thermally disordered material used in \Cref{fig:lof_nosoc_ws} do not; we can conclude that inversion symmetry does not play an essential role in the quenching of OAM currents. 

An unphysically large value of $B=40 \,$Ry in \eqref{eq3} was used to achieve the high degree of orbital polarization seen in \Cref{fig:lof} (where the Fermi energy in the lead was shifted to maximize it) and was chosen for illustrative purposes. The short length scale of the decay is unchanged on using a more reasonable value of $B=0.05 \,$Ry as in \Cref{fig:lof_ws}.
To test the effect of the orbital Zeeman splitting in the left lead, we calculated the orbital current injection for a number of values of $B$ in Pt. The orbital polarization achievable with these values of $|B|$ is shown in the inset to \Cref{fig:lof_ws}. For all polarization values, the orbital current decays to less that $1/e$ times its original value within three atomic layers.

\begin{figure}[t]
\centering
\includegraphics[width=8.4cm]{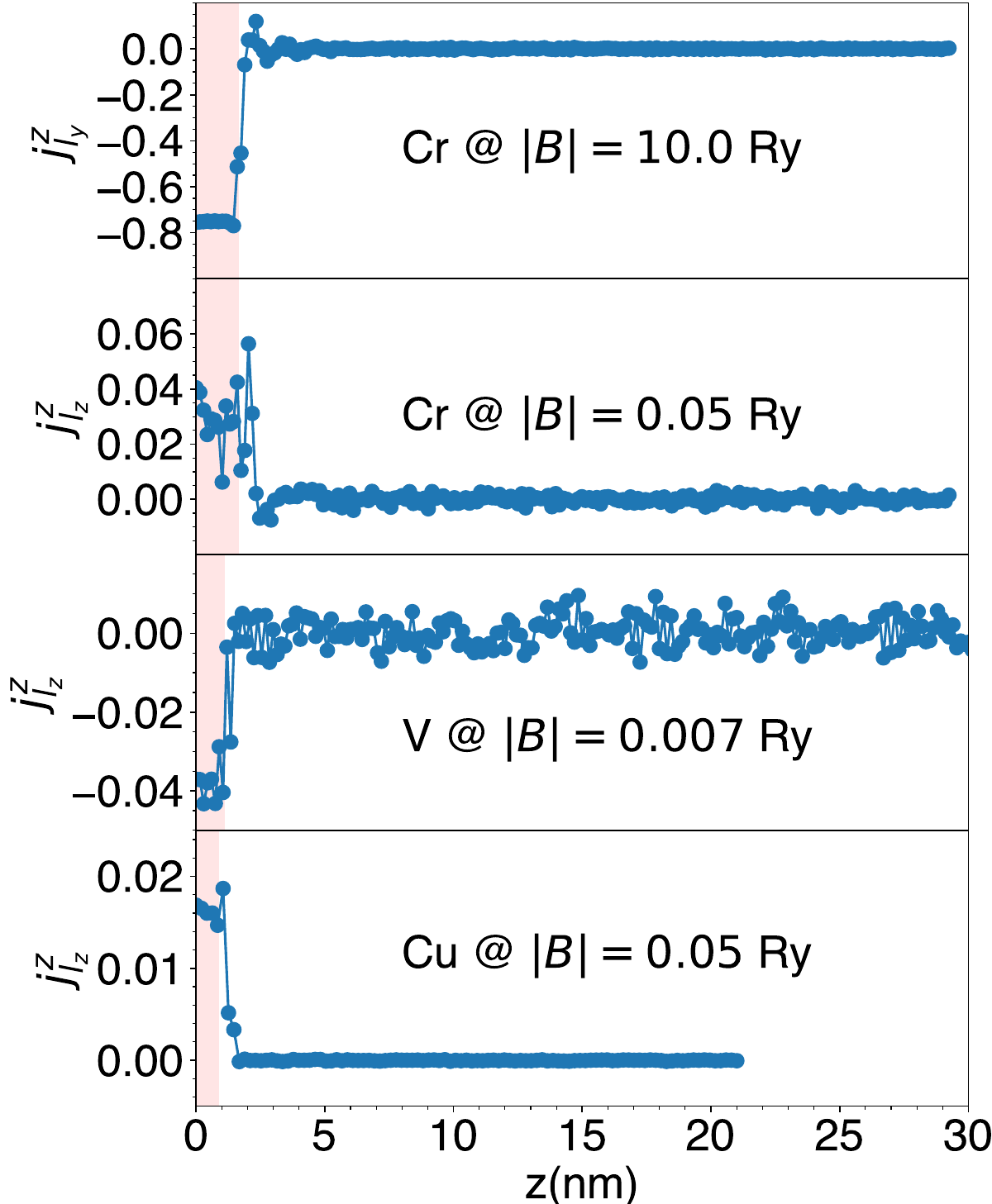}
\caption{The orbitally-polarized current injected from an orbitally-polarized left lead decays on a length scale of only a few atomic layers in thermally disordered Cr, V and Cu. For Cr, orbitally-polarizing fields have been used in the transport direction ($j_{lz}^z$) and perpendicular to it ($j_{ly}^z$). The orbital relaxation length is very short in all cases.}
\label{fig:lof_mult}
\end{figure}

As shown in \Cref{fig:lof_mult}, we observe a short decay length not only for Pt, but also for Cr, V and Cu. Although different amplitudes of $B$ can produce very different values of the orbital polarization, the orbital current decays to values indistinguishable from the fluctuations arising from disorder sampling within a few atomic layers in all cases. This indicates that the length scale $l_{\rm of}$ does not depend on the orbital Hall conductivity which we find to be large for Cr and V but small for Cu \cite{Salemi:prm22}.

Even a magnetic field as small as $B = 0.007$ Ry ($\approx 100$ meV) in \eqref{eq3} polarizes the lead sufficiently to allow the rapid decay of the orbital current injected into V to be clearly registered. In this case, we expect the electronic structure mismatch to be minimal. For the intermediate values of $B = 0.05$ Ry for Cu and $B = 10$ Ry for Cr we find the same behaviour; in all cases the orbital current is very short-ranged.
As in Pt, the orbital current injected into Cr and V generates a spin current through the spin-orbit coupling. However, the conversion rates in Cr and V are so low that the spin currents generated by the orbital-to-spin conversion are practically indistinguishable from the noise inherent in the disordered calculations.

Fitting an exponential curve to extract values for $l_{\rm of}$ for the different materials is not possible. We obtain rough estimates by counting the number of layers over which the polarization drops to $1/e$ times its value in the lead. 
As shown in \Cref{tbl:lof}, all $l_{\rm of}$ values determined in this way are $<1$ nm. The range shown for Cr follows from the slight difference between $y$ and $z$ orbital polarizations; in both cases, the value of $l_{\rm of}$ is very small.

\begin{table}
\caption{$\lof$ values for Pt, Cu, V and Cr extracted from the decay of orbital polarization inside the scattering region. The number of atomic layers after which the polarization is below the value of $1/e$ times the value in the lead is counted and multiplied by the atomic interlayer distance to obtain an upper bound for $\lof$.}
\begin{ruledtabular} 
\begin{tabular} {l | c | c | r}
   & $1/e$ at layer \# & $d$ (nm) per atomic layer & $\lof$ (nm)\\
\hline
Pt &    3 & 0.23 & 0.6 \\
Cu &    2 & 0.21 & 0.4 \\
 V &    3 & 0.14 & 0.5 \\
Cr & 2--5 & 0.15 & 0.3--0.8 \\
\end{tabular}
\end{ruledtabular}
\label{tbl:lof}
\end{table}

\section{Discussion}
Experiments measuring the orbital Hall effect have been interpreted in terms of long orbital diffusion lengths whereas the results presented here show that the polarization of an orbital current injected into a bulk metal drops off within a few atomic layers. The results for Pt show that the orbital current is converted into a spin current on the same short length scale. One possible explanation of the long length scale extracted from experiment is that what is being observed is the long $l_{\rm sf}$ over which the spin current resulting from the injected orbital current decays rather than the orbital current itself.

Exactly how the orbital current might be converted into a spin current in systems containing the light elements Cr and Ti is unclear. One possible explanation is that there is some orbital-to-spin conversion taking place at an interface or surface facilitated by the reduced symmetry of the interface. 
Another possibility is that a large spin-Hall current may be generated directly at the interface where the $3d$ materials simply serve to facilitate symmetry breaking by being different from the ferromagnetic material with which they are interfaced \cite {WangL:prl16}. 
Lastly, we note that large spin Hall angles are theoretically predicted in late 3$d$ metals where the reduced 3$d$ bandwidth compensates for the reduced SOC \cite{Gupta:19, Amin:prb19, Nair:21, Rang:24}; this is expected to greatly complicate the interpretation of experiments performed on bilayers \cite{Rang:24}

Though the precise mechanism may be dependent on the experiment being performed, it is well known that the effect of SOC is greatly enhanced by the symmetry lowering at an interface. 
The results presented here show that the orbital current can be converted into a spin current within a few atomic layers and this process could be compatible with an interface-mediated mechanism. 
One conclusion we can draw from the results presented here is that only a few atomic layers are needed to convert almost all of the orbital current into a spin current and a thick slab of orbital-to-spin conversion material may not be needed.

Our finding that $l_{\rm of}$ is very short is not necessarily bad news. The origin of the torque exerted by injecting a spin or orbital current from a nonmagnetic into a magnetic material is largely a matter of interpretation; for example, no change has to be made to electronic structure based computer codes that calculate spin-orbit torque (SOT) to include the OHE; it is included automatically. The important insight from OHE studies is that the search for efficient SOT need not be restricted to heavy metals with large spin-orbit coupling; the reduced bandwidth and frequently large state density at the Fermi energy of 3$d$ elements being frequently more important \cite{Gupta:19, Amin:prb19, Nair:21, Rang:24}. One consequence of a very short $l_{\rm of}$ is that the orbital-Hall conductivity is not a useful figure of merit in the search for better SOT materials. Another is that because the SOT is essentially a property of an A$|$B interface, it cannot be factored into separate A and B parts.

\section{Conclusion}
By computing the decay of orbital currents injected from an orbitally polarized lead, we have shown that the orbital relaxation length in transition metals is as short as might be expected from classical orbital quenching arguments. In all cases studied here, the orbital polarization decays to less than $1/e$ times its injected value within less than a nanometer. It does not matter if the polarization is small or large. The decay length does not depend on the spin Hall angle ($l_{\rm of}$ is short for both Pt with a large SHA and for Cu with a negligible SHA) or the orbital Hall angle ($l_{\rm of}$ is short for V and Cr as well as for Cu). The conversion of the orbital current into a spin current does significantly depend on the spin Hall angle, as expected. 
These results contradict the interpretation of experimental findings in terms of  significantly longer length scales. One possible explanation is that the experiments are actually measuring the spin-flip diffusion length instead. In the case of Pt, we have shown how the injected orbital current is converted into a spin current within a few layers after which it decays as we expect for the spin current.

\section{Acknowledgement}
This work was sponsored by NWO Domain Science for the use of supercomputer facilities.
We acknowledge discussing the role of thermal disorder with Sergei Urazhdin and Thierry Valet.


\begin{thebibliography}{50}
\bibitem [{foo({\natexlab{a}})}]{footnote2}%
  \BibitemOpen
  \href@noop {} \bibinfo {note} {{Because of how currents
  are extracted from the scattering wave functions \cite{Wesselink:prb19},
  $j(z)$ is not calculated in the leads. To visualise $j(z)$ in the left-hand
  lead ($\mathcal{L}$), the $\mathcal{L|S}$ interface is artificially shifted
  into the scattering region to $z$=$z_I$. The region $z<z_I$ where the
  magnetic field $\hat{B}$ in \eqref{eq2} and \eqref{eq3} is nonzero and there
  is no thermal disorder is indicated in the figures by pink shading. For
  simplicity, this effective lead will be referred to as the lead and the part
  of the scattering region $z>z_I$ will be referred to as the scattering
  region}}\BibitemShut {NoStop}%
\bibitem [{foo({\natexlab{b}})}]{footnote1}%
  \BibitemOpen
  \href@noop {} \bibinfo {note} {{For convenience, we
  measure the charge density in units of the electron charge -$e$ (where $e$ is
  a positive quantity), the spin density in units of $\hbar / 2$ and the
  orbital density in units of $\hbar$ so they all become particle densities and
  the fluxes become particle fluxes.}}\BibitemShut {Stop}%
\bibitem [{foo({\natexlab{c}})}]{footnote3}%
  \BibitemOpen
  \href@noop {} \bibinfo {note} {{Our interpretation of
  the very small value of $l_{\rm of}$ in terms of orbital quenching and time
  reversal symmetry becomes more complicated at $T=0 \,$K when orbital
  degeneracies occur. Nevertheless, the numerical calculations are easily
  performed in the absence of thermal disorder and $l_{\rm of}$ does not change
  appreciably. Thus thermal disorder is not responsible for the quenching of
  orbital currents.}}\BibitemShut {Stop}%
\end{thebibliography}
\end{document}